\documentclass[useAMS,usenatbib]{mn2e}
\usepackage{graphicx}
\usepackage{longtable}
\usepackage{amssymb,amsmath}

\newcommand{\msol}{$M_{\odot}$}

\newcommand{\wcen}{$\omega$~Cen}
\title
{On the nature of the brightest globular cluster in M81}

\author[Mayya et al.]{Y.D. Mayya$^{1}$\thanks{Email: ydm@inaoep.mx},
D. Rosa-Gonz\'alez$^{1}$, 
M. Santiago-Cort\'es$^{1}$, 
L.H. Rodr{\'\i}guez-Merino$^{1}$, 
\newauthor
O. Vega$^{1}$, 
J.P. Torres-Papaqui$^{2}$, 
A. Bressan$^{3}$
and 
L. Carrasco$^{1}$ \\
$^{1}$Instituto Nacional de Astrof\'isica \'Optica y Electr\'onica, Luis Enrique Erro 1, Tonantzintla 72840, Puebla, Mexico \\
$^{2}$Departamento de Astronom\'{\i}a, Universidad de Guanajuato, Apartado Postal 144, 36000, Guanajuato, Mexico \\
$^{3}$SISSA, via Bonomea 265, 34136, Trieste, Italy.
}

\begin{document}
\date{{\bf Accepted in MNRAS (September 2013)}}
\pagerange{\pageref{firstpage}--\pageref{lastpage}} \pubyear{2009}

\maketitle

\label{firstpage}

\begin{abstract}
We analyse the photometric, chemical, star formation history and structural 
properties of the brightest globular cluster (GC) in M81, referred as GC1 in 
this work, with the intention of establishing its nature and origin.
We find that it is a metal-rich ([Fe/H]=$-0.60\pm0.10$), alpha-enhanced 
([$\alpha/{\rm Fe}]\sim0.20\pm0.05$), core-collapsed (core radius $r_c=1.2$~pc, tidal
radius $r_t = 76r_c$), old ($>13$~Gyr) cluster. It has an ultraviolet excess 
equivalent of $\sim2500$ blue horizontal branch stars. It is detected in X-rays
indicative of the presence of low-mass binaries.
With a mass of $1.0\times10^7$~\msol, the cluster is comparable in mass to 
M31-G1 and is four times more massive than $\omega$~Cen. 
The values of $r_c$, absolute magnitude and mean surface brightness of GC1 
suggest that it could be, like massive GCs in other giant galaxies, the
left-over nucleus of a dissolved dwarf galaxy.
\end{abstract}

\begin{keywords}
catalogs -- galaxies: individual (M81) -- galaxies: spiral -- galaxies: star 
clusters --- globular clusters: general
\end{keywords}

\section{INTRODUCTION }
The most massive globular clusters (GCs) (mass $\ga2\times10^6$~\msol) in galaxies 
are found to be different from the rest of the GC population in their 
physical, chemical and dynamical properties
\citep{Hasegan2005, Mieske2006, Georgiev2009, Taylor2010, 2012Jang}. 
These massive clusters seem to be 
related to the higher mass compact systems such as 
nuclear clusters \citep{Georgiev2009} and ultra-compact dwarfs (UCDs) 
\citep{Phillipps2001} rather than to the lower-mass classical GCs. 
Thus, it seems unlikely that the massive GCs were formed {\it in-situ}
in the halos of their present parent galaxies, like their lower mass
counterparts. Meanwhile, there is growing evidence in support of
the idea proposed by \cite{Zinnecker1988}, 
that was later tested using numerical simulations by \cite{Bekki2003}, 
that some of the massive compact objects could be left-over nuclei of tidally
stripped dwarf galaxies.
Well-known examples of these type of clusters are \wcen\ in our galaxy 
\citep{Meylan1995}, Mayall II (G1) in M31 \citep{2001Meylan}, 
and several massive GCs in NGC5128 \citep{Taylor2010}.

The brightest GC in a galaxy would also be the most massive if it is old
like most GCs. GCs in elliptical galaxies all seem to be older than 10~Gyr 
\citep{2006Brodie, 2005Strader}, the exception being the gas-rich elliptical NGC1316 
which contains intermediate-age (3--10~Gyr) GCs \citep{2001Goudfrooij}. 
However, there is clear evidence of GC-like objects forming at present epochs
in gas-rich spirals that have experienced merging \citep[e.g. Antennae:][]{1995Whit}.
Even mild interactions are able to trigger the formation
of massive compact objects such as the case of M82, which has formed a 
population of compact clusters in its disk following 
its interaction with M81 around 500~Myr ago \citep{Mayya08,Konstan09}.
It is not yet established whether this interaction or a similar interaction 
in the past was able to form any massive compact clusters in M81, that would
have observational properties of old GCs.

In a systematic search for compact clusters in M81 using the HST/ACS images,
\cite{2010Santiago} identified R05R06584 (GC1 henceforth; RA=09:55:22.042 
$\delta=+$69:06:37.84 (J2000)) as the brightest among the 172 GCs in this galaxy.
It was noted in that work that the cluster is more luminous than the brightest 
GCs in the Milky Way ($\omega$~Cen) and Andromeda (M31-G1), the only two galaxies 
of comparable mass that are closer to us than M81. 
GC1 is seen at a projected galactocentric distance of only 3.0~kpc,
as compared to galactocentric distances of 6.3~kpc, and 40~kpc of 
$\omega$ Cen and M31-G1, respectively. 
The object had been previously identified as 50777 in \citet{PerelmuterR1995} 
and was the target of a spectroscopic study by \cite{2010Nantais} (their 
object 1029), who reported a metallicity of [Fe/H]$=-0.86\pm0.41$. 
They calculated this metallicity using emperical relations between Lick 
indices and metallicity \citep{Brodie1990}. 
They also reported a radial velocity of 
$131\pm5$~km\,s$^{-1}$, which is consistent with radial velocity
obtained for disk objects at the observed galacto-centric distance of GC1.

In spite of being the brightest GC in M81, the nature of GC1 is unknown.
If it is an old GC like $\omega$~Cen or M31-G1, it would be the most massive
GC not only in M81, but also in the local Universe 
(distance $\lesssim3.6$~Mpc). On the other hand, if it was
formed later on, its mass would be smaller.
Determination of its age is critical to distinguish these two possibilities. 
The cluster is not resolved into
individual stars even on the HST/ACS images, and hence age cannot be 
obtained using the classical technique of colour-magnitude diagram (CMD).
In this paper, we determine the age using two independent methods: 
(1) by fitting the optical spectrum between 3600--7000~\AA\ with a model
spectrum that is constructed as a sum of several Single Stellar
Populations (SSPs) using the code {\scriptsize STARLIGHT}, and
(2) by fitting the UV to MIR spectral energy distribution (SED)
with SSPs of fixed ages. The latter method is very powerful in inferring the
presence of intermediate-age populations \citep{Bressan1998}. 
In Table~\ref{Tab:SuperClusters}, we compare the absolute magnitude ($M_V$), 
metallicity ([Fe/H]), $V-I$ colour at the cluster center, age, mass, central
$V$-band surface brightness ($\mu(0,V)$), and the structural parameters of 
GC1 to the corresponding parameters in $\omega$~Cen and M31-G1.
The tabulated masses are photometric masses. 
Two additional values of masses are given for M31-G1 inside parentheses, 
both obtained using dynamical methods by \cite{2001Meylan}, the smaller one 
is the Virial mass and the larger one is the King mass.

In \S2, we describe observational data used in this work. The method we
adopted to obtain the structural properties is given in \S3. In \S4, we 
describe the analysis technique we have used to obtain the metallicity,
age, extinction and mass of GC1. The nature and the origin of GC1 are discussed in \S5.

\begin{table*}
\begin{center}
\caption{\label{Tab:SuperClusters} Structural and physical properties of M81-GC1 compared to those in 
$\omega$ Cen and M31-G1.}
\begin{tabular}{lccccccccccc}\hline
Name         &  Mv(GC) & [Fe/H] & $(V-I)_0$ & Age &  Mass & $\mu(0,V)$ & $c=\log(r_t/r_c)$ & r$_c$& r$_h$ & $\epsilon$ & Ref.$^1$       \\ 
 \           &  mag    &     & mag   &  Gyr& $10^6$~\msol\ & mag arcsec$^{-2}$  &      &  pc  & pc   &\  & \  \\\hline
$\omega$ Cen        & $-10.37$ & $-1.62$ & 0.88 & $>13$ & 2.3 & 14.38 & 0.98 & 3.58 & 7.71 & 0.17 &  (a,b)     \\
M31-G1 (Mayall II)  & $-10.90$ & $-0.95$ & 1.10 & $>13$ & 8.6(7.3,15)  & 13.47 & 2.50 & 0.52 & 13.5 & 0.20 &  (c,d)     \\ 
M81-GC1 (R05R06584) & $-11.40$ & $-0.60$ & 0.96 & $>13$ & 10  & 14.93 & 1.88 & 1.23 & 5.60 & 0.12&   (e,f)  \\
\hline
\end{tabular}\\
$^1$ References are (a)\cite{1996Harris}, (b) \cite{Georgiev2009}, (c)\cite{2001Meylan}, 
(d) \cite{2009Ma}, (e) \cite{2010Santiago} and (f) this paper.
\end{center}
\end{table*}

\section{OBSERVATIONAL DATA}

\subsection{HST Imaging data }

\begin{figure}
\scalebox{.45}{\includegraphics{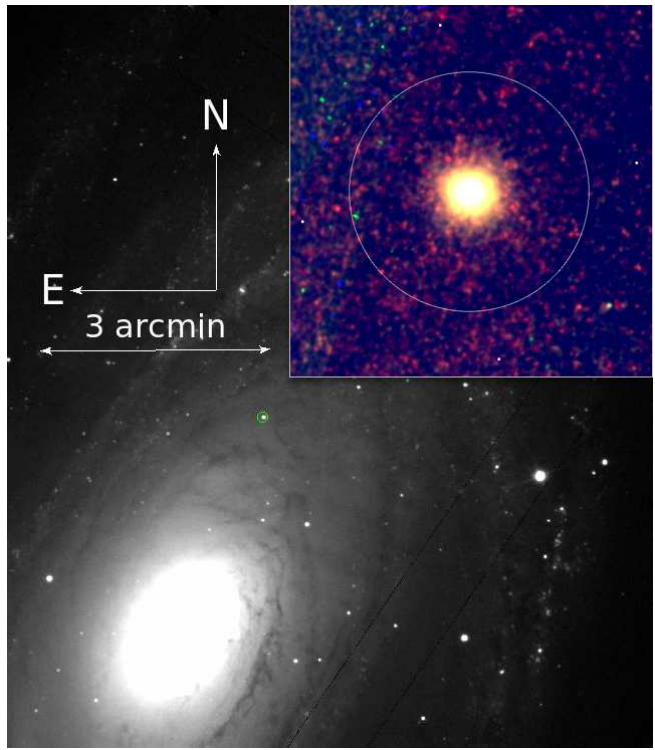}}
\caption{\label{fig:gc1rgbim} 
SDSS $g$-band image of M81 showing the location of GC1 with a circle.
The inset contains an RGB image formed from the HST F814W (red),
F606W (green) and F435W (blue) images. The circle around GC1 is of 4\arcsec\ 
(75~pc) radius, which is the aperture used for multi-band photometry.
}
\end{figure}

The imaging data that we used in this work to obtain the structural parameters
come from the  \textit{HST} observations in the F435W, F606W and F814W filters 
(PI: Andreas Zezas). These images have a sampling
of $0\farcs05~{\rm pix}^{-1}$ which at the distance of 3.63~Mpc 
to M81 \citep{Freedman1994} corresponds to 0.88~pc~${\rm pix}^{-1}$.
The spatial resolution in these images, measured as the Full Width at Half 
Maximum of the Point Spread Function (PSF), is 2.1~pixel (1.8~pc).
The cluster is located at a projected distance of 3~kpc from the nucleus
within 10\degr\ of the north-west major axis of M81, as illustrated in 
Figure~\ref{fig:gc1rgbim}. A blow-up of GC1 in a colour-composite HST image 
is also shown in this figure.

\subsection{Spectroscopic observations }

Spectroscopic observations were carried out using
the long-slit of the spectrograph of the OSIRIS instrument at the 10.4-m
GTC\footnote{Gran Telescopio Canarias is a Spanish initiative with the
participation of Mexico and the US University of Florida,
and is installed at the Roque de los Muchachos in the island of La Palma.
This work is based on the proposal GTC11-10AMEX 0001.} in the service mode on 
2010 April 6. The observations included bias, flat fields, calibration 
lamps and standard star. A slit-width of 1\arcsec\ was used and 3 exposures 
of 900 seconds each were realized, using the R1000B grism. The $2\times1$ 
detector binning gives a spatial scale of 0.25 arcsec pix$^{-1}$, and
spectral resolution of $\sim$7~\AA\  at 5510~\AA.
The seeing during the observations
was $\sim1\arcsec$. 

The data reduction was carried out in the standard manner using the
tasks available in the IRAF\footnote{IRAF is distributed by the National Optical
Astronomy Observatory, which is operated by the Association
of Universities for Research in Astronomy (AURA) under
cooperative agreement with the National Science Foundation.} 
software package. The individual spectra 
were debiased, flat field and illumination corrected, wavelength 
calibrated and background subtracted. At the end of the reduction 
procedure, we combined the 3 
different spectra and obtained a final spectrum free of cosmic rays.
Observations of Feige 34 during the same night were used for flux calibration.

\subsection{Multi-band photometric data }

M81 was a target of the Spitzer Infrared Nearby Galaxies Survey (SINGS:
\cite{Kennicutt2003,Perez2006}). In addition to these mid infrared
images, the galaxy was part of surveys at ultraviolet (Galex), optical 
(Sloan Digital Sky Survey; SDSS), and near infrared (2MASS) wavelengths. 
Fits format files from these missions are available at NED\footnote{This 
research has made use of the NASA/IPAC Extragalactic Database (NED)
 which is operated by the Jet Propulsion Laboratory, California Institute
 of Technology, under contract with the
 National Aeronautics and Space Administration.}, which were used 
in this study.

\subsubsection{Aperture photometry }
We carried out aperture photometry on archival images in two Galex bands, five 
SDSS filters, the 2MASS JHK bands, the four bands of Spitzer/IRAC and the
24~$\mu$m band of Spitzer/MIPS. The object is clearly detected in all of these
bands, except the 24~$\mu$m band, where we determined an upper limit. The 
relative isolation of the object allowed us to use an aperture of radius of 
4\arcsec (75~pc) that is big enough to include more than 80\% of the total 
flux in most bands. The aperture fluxes were multiplied by correction factors 
to account for the flux outside the aperture to obtain the total flux of GC1. 
The correction factor was more than 20\% for the following three bands: 1.72 
for Galex (NUV), 1.39 for Galex (FUV), and 1.22 for the 8~$\mu$m band of 
Spitzer/IRAC. For obtaining the aperture magnitudes, the sky was chosen in 
an annular region of 3\arcsec\ width starting at 7\arcsec\ radius. 
The instrumental magnitudes are converted into the system of ABmag
(and Jansky) using the calibration constants in the headers of the respective images.
Errors ($\delta F$) on the measured fluxes ($F$) are calculated as 
\begin{equation}
{\delta F \over{F}} = {\sqrt{F + N\sigma_{\rm sky}^2 T} \over{F\sqrt{T}}},
\end{equation}
where $F$ is the count rate of photons measured in the aperture, $N$ is the 
number of pixels in the aperture, $\sigma_{\rm sky}$ the sky rms/second/pixel as measured 
in the part of the image outside the main galaxy, and $T$, the total exposure
time in seconds in each image.

The multi-band integrated fluxes in ABMAGs and Janskys, along with their errors are 
given in Table~2.

\begin{table}
\begin{center}
\caption{\label{Tab:photometry} Multi-band integrated fluxes of GC1}
\begin{tabular}{llccc}\hline
$\lambda$ & Mission/Filter & ABMAG & $F_\nu$ & $\delta F_\nu/F_\nu$ \\ 
\AA &  & mag & Jy &  \\ \hline
1528   & Galex-FUV   &  21.527 &   1.2367E-5 &  0.066 \\
2271   & Galex-NUV   &  21.239 &   1.9955E-5 &  0.034 \\
3551   & SDSS-u      &  18.861 &  1.0778E-4  &  0.023 \\
4686   & SDSS-g      &  17.257 &   4.5854E-4 &  0.002 \\
6165   & SDSS-r      &  16.551 &   8.9660E-4 &  0.002 \\
7481   & SDSS-i      &  16.158 &   1.2742E-3 &  0.001 \\
8931   & SDSS-z      &  15.846 &   1.7161E-3 &  0.005 \\
12350  & 2MASS-J     &  15.595 &  2.2043E-3  &  0.024 \\
16620  & 2MASS-H     &  15.480 &  2.4726E-3  &  0.027 \\
21590  & 2MASS-K     &  15.820 &  1.7748E-3  &  0.034 \\
35500  & Spitzer-3.6 &  16.573 &  9.4611E-4  &  0.002 \\
44900  & Spitzer-4.5 &  17.081 &  5.9266E-4  &  0.003 \\
57300  & Spitzer-5.8 &  17.478 &  4.1880E-4  &  0.016 \\
78700  & Spitzer-8.0 &  18.367 &  1.9936E-4  &  0.037 \\
236800 & Spitzer-24  &  $>$20.570 &  $<$2.1478E-5  &  0.300 \\
\hline
\end{tabular}\\
\end{center}
\end{table}

\section{Determination of Structural Parameters }

We used the IRAF/STSDAS software package {\it ellipse} to analyse the 
structural parameters of GC1 in the HST images. {\it Ellipse} is an 
algorithm designed to fit isophotes of galaxies with ellipses, where the intensity 
profiles decrease monotonically with radius.
The first requirement of the analysis is to establish the best
 photometric centroid for the cluster. We started with the reported position
of GC1 in \cite{2010Santiago}, which is the centroid defined by 
SExtractor \citep{Bertin1996}. We refined this centre by re-calculating the
centre as the average centre of ellipses with semi-major axis $<5$~pixels.
The resulting centre differed from that reported by \cite{2010Santiago}
by less than 1 pixel. However, we obtained surface brightness profiles
by fixing the centre at this newly obtained value.
Other quantities that affect the analysis of structural parameters are
the choice of the background level and the range of radius used for 
fitting. We estimated the local background around GC1 in small ($10\times10$~pixel$^2$) 
boxes, and defined the fitting radius as the semi-major axis where the
azimuthally averaged intensity reaches the previously measured background level.
The cluster is devoid of any major contaminating sources in its immediate 
vicinity within 3\arcsec\ radius that is considered in the analysis of the
profiles --- the nearest contaminating star is of $B=25.7$~mag (VEGAMAG) at 
a radial distance of 3\farcs26 (66 pixels) from the cluster centre. 
Finally, we subtracted the background value from the azimuthally averaged 
intensity profiles to obtain sky subtracted profiles in $F435W$ and $F814W$ bands.

\begin{figure}
\scalebox{0.6}{\includegraphics{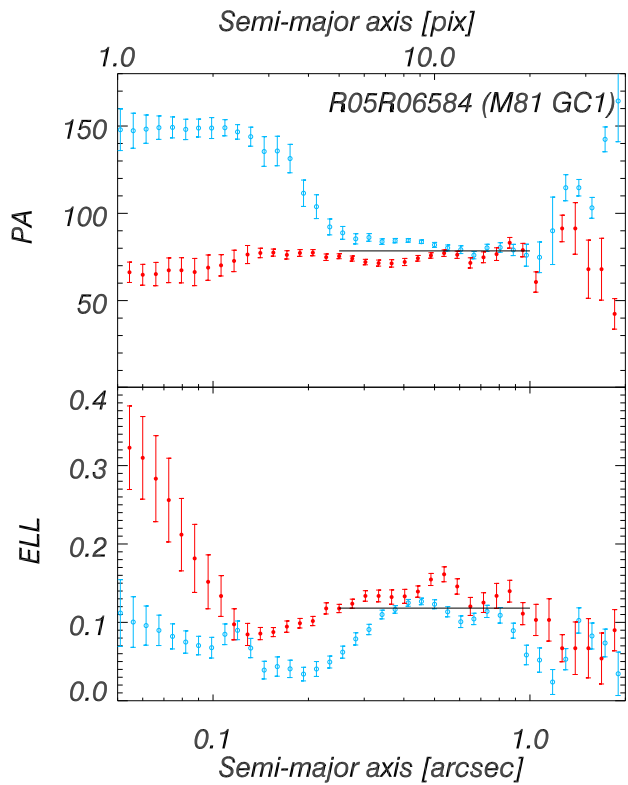}}
\caption{\label{fig:gc1pa}
Variation of the ellipticity (bottom) and position angle (PA; measured
from north to east) of the major axis (top) of the ellipses that best fit the
isophotal intensities in $F435W$ (blue symbols) and $F814W$ (red symbols) bands.
The values in both the bands remain constant between 5--20~pixels at values of 
PA=$78.5\pm5$\degr\ and $\epsilon=0.12\pm0.02$, which are shown by the 
horizontal bars in the corresponding panels.
}
\end{figure}

 We also  analysed the variation of ellipticity ($\epsilon=1-\frac{b}{a}$)
 and the position angle of the major axis (PA) of the cluster for increasingly
 larger ellipses of fixed centres. The 
variation of $\epsilon$ and the PA as a function of semi-major axis length
in $F435W$ and $F814W$ bands are shown in Figure~\ref{fig:gc1pa}. 
The PA and $\epsilon$ in the two filters remain almost constant between 
5--20~pixels, at values of PA=$78.5\pm5$\degr\ and $\epsilon=0.12\pm0.02$.
The values in the inner-most two pixels are limited by resolution
 and hence the observed differences in the two bands are not significant.
The cluster is elongated almost along the perpendicular direction to the 
semi-major axis of the parent galaxy.

The dynamical history of star clusters can be investigated through the 
analysis of their surface 
brightness profiles. It is well known that surface brightness profiles of 
most GCs in the Milky Way, M31 and M33 are reasonably well represented by the King 
model profile \citep{King1962,McLaughlin2005}. Consequently, we fit the radial
intensity profile of GC1 with an empirical King model 
\citep{King1962,King1966} after convolving
it with the PSF of the image in each filter. The characteristic 
PSF was derived using the photometry routine IRAF/DAOPHOT using selected 
isolated stars uniformly distributed over the entire HST images that contain the GC1 
star cluster in $F435W$ and $F814W$ bands.
The procedure we followed to realize the fits is identical to that in ISHAPE
\citep{1999Larsen}, except that our procedure allows the determination of 
the tidal radius $r_t$ directly from the fits.

\begin{figure}
\scalebox{1.0}{\includegraphics{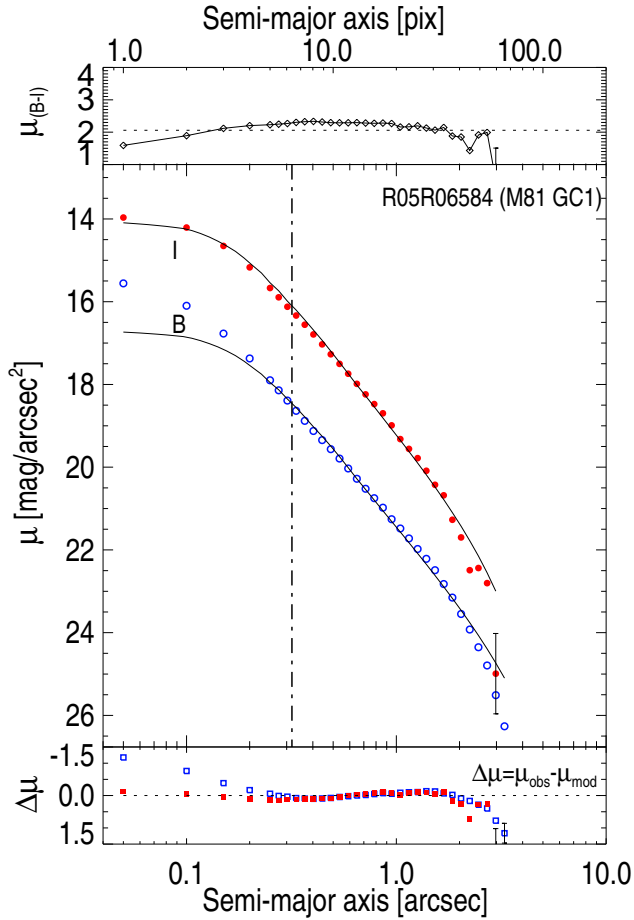}}
\caption{\label{fig:gc1prof}
In the middle panel, the observed azimuthally averaged surface brightness 
profile for GC1 in the $F435W$ (empty circles) and $F814W$ (filled circles) 
bands, along with the best-fitting King profiles (solid lines) are shown. 
The vertical line represents the effective radius in the $F814W$ band. 
In the bottom panel, we show the residual (observed$-$King)
surface brightness in $F435W$ (empty squares) and $F814W$ (filled squares) bands.
Only the error bars with values greater than the symbol size are plotted.
The $F435W - F814W$ colour profile is shown in the top panel, where
the horizontal dashed line corresponds to the integrated colour of the
cluster.
}
\end{figure}

The observed surface brightness profiles of CG1 in $F435W$ and $F814W$ bands 
are shown in Figure~\ref{fig:gc1prof}. Superposed on these observed profiles 
are the best fitting King model profiles (after convolving with the image 
PSFs). These models have core radius $r_c=1.80$~pix and tidal radius $r_t=146$~pix 
for the $F435W$ band, and $r_c=1.40$~pix and $r_t=106$~pix for the $F814W$ band. 
Whereas the King profile fits very well the observed profile over the entire plotted
range in the $F814W$ band, the observed $F435W$-band central surface brightness 
is $\sim1.0$~mag brighter than that for the best-fit King profile. This apparent "blue core" 
is also seen in the colour profile (top panel), which is most likely related 
to the presence of blue horizontal branch stars that also produces an UV excess 
as discussed in \S4.3. Thus, it is advisable not to use the blue-band profiles
for obtaining structural parameters. Hence, we used the $r_c=1.40$~pix (1.2~pc) and 
$r_t=106$~pix (93~pc) for the $F814W$-band as typical values for GC1.
This results in a value of $r_t/r_c=76$.

The {\it ellipse} task also performs photometry in successive ellipses.
We used these photometric data to determine $R_{\rm eff}$, the radius
where the aperture flux is half the total flux (also known as half-light
radius $r_h$), in each of the two bands. 
The $R_{\rm eff}$ in the $F814W$ band is 5.6~pc.
We also fitted a Moffatt model profile in both the filters. 
The King model fits the profiles better, especially in the outer parts.

\section{Determination of Physical parameters}

\subsection{New determination of [Fe/H] and [$\alpha$/Fe]}

\begin{figure}
\scalebox{0.50}{\includegraphics{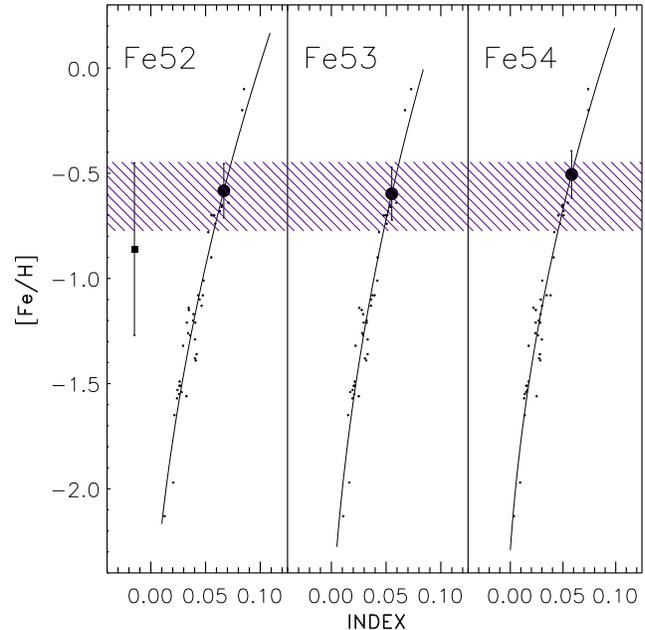}}
\caption{\label{fig:metal} 
An illustration of the method used to determine [Fe/H] of GC1 from the 3 iron 
indices. The values for the Galactic GCs of Schiavon et al. (2005) (dots)
are fitted with a polynomial of second order (solid line) to obtain a new emperical
calibration. Observed values of indices for GC1
are plotted at their [Fe/H] values inferred from our emperical calibration, with the shaded
area denoting our mean value of [Fe/H]$=-0.60\pm0.10$. 
The [Fe/H] reported by Nantais \& Huchra (2010)
is shown by the filled square along with their error, on the left most
panel. Our calibration not only has reduced the errors on the measurement, but also
results in 0.26~dex higher metallicity. 
}
\end{figure}

\cite{2010Nantais} derived a metallicity [Fe/H]$=-0.86\pm0.41$ for GC1 using
empirical calibration of the indices defined by \cite{Brodie1990} and \cite{Trager1998}
for Galactic GCs \citep{Schiavon2005}. Their value is the weighted
mean of [Fe/H] values derived using indices MgH, Mg2, Mgb, F25270, Fe5335, Fe5406, G4300,  $\delta$
and CNR, with more weights given to indices with larger dynamic range, defined as
the ratio of sensitivity of the index (a change of 1 dex in [Fe/H]) to the observational
error of the index. Such a weighting scheme doesn't foresee variations
of [$\alpha/Fe$] in GCs and hence would give erroneous values of [Fe/H] for systems that
have [$\alpha/Fe$] values different from the mean value for the Galactic sample.
The relatively large error in their measured value is an indicator of dispersion
in the measured abundances using different indices. 
Low signal-to-noise ratio of their spectra may also be responsible for the large scatter.

We used our GTC spectra of GC1 to improve the value of [Fe/H] and also to newly determine the
[$\alpha/Fe$]. We measured the indices Fe5270, Fe5335 and Fe5406 and determined the [Fe/H]
as the mean of metallicities obtained from each of these indices as is illustrated in
Figure~\ref{fig:metal}.
The [Fe/H] vs index data for the Galactic GCs of \cite{Schiavon2005} (dots)
is fitted with a polynomial of second order (solid line) to obtain a new empirical
calibration. Note that the fits clearly illustrate the quadratic nature of the
relation, though it is a common practice to fit these points with a straight line 
\citep[e.g.][]{2010Nantais}.
Observed values of indices for GC1
are plotted at the [Fe/H] values inferred from our empirical calibration, with the shaded 
area denoting our mean value of [Fe/H]$=-0.60\pm0.10$. 
The [Fe/H] reported by 
\cite{2010Nantais} is shown by the filled square along with their error on the left most
panel. Their large error bar is due to the use of non-iron indices to measure [Fe/H], and
also the use of linear fits between indices and [Fe/H]. If we use our calibration of the 
three iron indices with the index values reported by them (Table 4 \& Table 5 in their paper), 
we obtain [Fe/H]$=-0.48\pm0.17$ which is in agreement with values from our spectra.

We also used the spectrum of GC1 to measure the Lick indices ratio Mgb/$<$Fe$>$ 
\citep{Worthey1994}, where $<$Fe$>=(Fe5270+Fe5335)/2.0$, and obtained the 
$\alpha$ element enhancement using the relation of \cite{Thomas2003}. 
This gives us a value of $[\alpha/Fe]=0.20\pm0.05$.
Adopting the relation of \cite{Annibali2007} between the indices and 
metallicity, our observed values of [Fe/H] and $[\alpha/Fe]$ correspond 
to $Z=0.0056$ for the commonly used value of $Z\odot=0.02$. However,
for the recently revised value of $Z\odot=0.014$ \citep{Asplund2009}, 
we get $Z=0.0043$. 

\subsection{Age of the Cluster from optical spectrum}

\begin{figure*}
\scalebox{.7}{\includegraphics{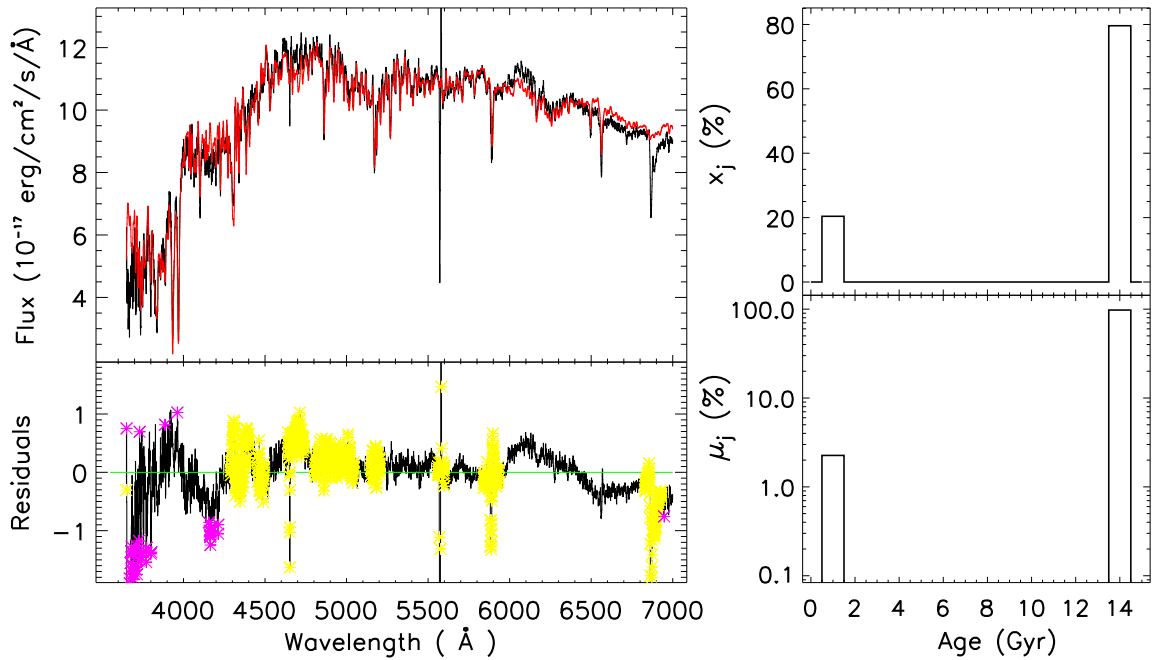}}
\caption{\label{fig:StarFit} Results from the STARLIGHT analysis. The top left
panel shows the best-fit model spectrum (red) overplotted on the observed spectrum 
(black). The residual spectrum in units of $10^{-17}$~erg\,cm$^2$\,s$^{-1}$\,\AA$^{-1}$
is ploted at the bottom left panel, where the masked out regions are shown
by yellow bands (emission lines) and pink asterics (bad pixels). 
The panels on the right show 
the star formation history of the cluster: top, the percentage in flux of
different stellar populations at the reference wavelength (4020~\AA), 
and the bottom, the percentage in mass, both as a function of age of the population. 
The stars that are formed $\gtrsim13$~Gyr ago account 
for $\sim 98$\% of the mass of the cluster.
}
\end{figure*}

We analysed our GTC spectrum of GC1 to determine the age of the stellar 
population, using the  \begin{scriptsize}STARLIGHT\end{scriptsize}
spectral synthesis code \citep{2005Cid, 2006Mateus}.
As a first step, we corrected the observed spectrum for Galactic extinction of
$Av= 0.22$ (maps of \cite{1998Schlegel} and the reddening  curve of 
\cite{Fitzpatrick1999}, using $R_V = 3.1$).
The spectrum was then brought to the rest frame wavelength, and also was 
resampled to pixels of 1\AA\ between 3600 and 7000\AA.
The \begin{scriptsize}STARLIGHT\end{scriptsize} decomposes an observed
spectrum in to a number of simple stellar populations (SSPs), each of
which contributes a fraction $x_j$ to the flux at a chosen normalization
wavelength (in our case $\lambda_0$ = 4020~\AA). We used 24 SSPs at a fixed metallicity 
of $Z=0.008$, the value closest to that observed in GC1 for which SSPs are
available in \begin{scriptsize}STARLIGHT\end{scriptsize}). 
The SSPs were extracted from the models of \cite{2003Bruzual}, computed 
for a \cite{1955Salpeter} initial mass function (IMF), `‘{\it Padova-1994}’' 
evolutionary  tracks \citep{1993Alongi, 1993Bressan, 1994aFagotto,
1994bFagotto, Girardi1996}, and STELIB library of observed stellar spectra
\citep{2003LeBorgne}. 
The ages of these SSPs range from 1~Myr to
14 Gyr, at approximately logarithmic steps.
Bad pixels and emission lines are masked and left out of the fits. 

The results of the \begin{scriptsize}STARLIGHT\end{scriptsize} analysis
are shown in Figure~\ref{fig:StarFit}. All the characteristics of the observed
spectrum are very well reproduced by the fit, with the residuals well below
10\% in most parts of the spectrum. In spite of using SSPs of 24 ages, we find
nearly 98\% of the stellar mass corresponding to stars that formed at the very 
early epochs of galaxy formation (age $>13$~Gyr) with an age spread $<2$~Gyr. 
The best-fit model
suggests that around 20\% of the blue light comes from another population
--- 1~Gyr old population of $\sim2$\% of total mass. In the next section, using
the fits to the entire SED, we will show that the source of this blue excess 
is most likely, the extreme blue horizontal branch stars, that are
not taken into account in the base SSPs in \begin{scriptsize}STARLIGHT\end{scriptsize},
rather than a 1~Gyr old population. Thus, the optical spectrum of GC1 is 
consistent with an age of $\gtrsim$13~Gyr.

\subsection{Age, Metallicity, Extinction and Cluster Mass from SED Analysis }

The effects of the presence of dusty circumstellar envelopes around asymptotic giant branch
(AGB) stars appear at wavelengths longward of a few microns and leave
 a clear excess around $10-15$ $\mu$m in the integrated
 mid-infrared (MIR) spectrum of passively evolving
systems \citep{Bregman1998, Bressan1998}. Since
AGB stars are luminous tracers of intermediate-age
stellar populations, the presence or not of their characteristic MIR excess
has been suggested as a powerful method to disentangle age and
metallicity effects among these systems \citep{Bressan1998, Bressan2001, Bressan2006}.
More specifically, the analysis of SSP models accounting
for the effects of dusty AGB stars \citep{Bressan1998}
shows that a degeneracy between metallicity and age persists even in
the MIR, since both age and metallicity affect mass-loss and
evolutionary lifetimes on the AGB. While in the optical regime, age
and metallicity need to be anti-correlated to maintain a feature
unchanged (either colour or narrow-band index), in the MIR it
is the opposite: the larger mass loss of a higher metallicity
simple stellar population (SSP) must be balanced by its older
age. Therefore, the detailed comparison of the MIR and optical data
of passively evolving systems constitutes perhaps one of the
cleanest ways to remove the degeneracy. The third parameter in the problem
of the degeneracy is the extinction. In recent years, several studies
 \citep[e.g.][]{Bianchi2005, Kaviraj2007a, Bridzius2008, Rodriguez-Merino2011}
have shown that the analysis of photometric
data from the UV to the NIR spectral range help to disentangle the effects of
reddening from those of evolution. However, they note that the derived
metallicities do not reach the accuracies achievable by using spectroscopic data.

In this section, we follow  the approach of the analysis of the panchromatic
SED, with the innovation of having a wider spectral range, from the UV to the MIR,
and by comparing these data with suitable SSP models accounting for the
effects of dusty AGB stars \citep{Bressan1998, Bressan2012}.
Notice that the photometric data reported in Table~2 correspond to fluxes
integrated over the entire cluster, which ensures reliable
results from the analysis of SED.

As a first step, we corrected the observed SED for Galactic extinction of
$Av= 0.22$, and the reddening  curve of \cite{Fitzpatrick1999} and \cite{Schlafly2011}.
The library of SSP model spectra were computed using the latest
release of the PARSEC evolutionary tracks \citep{Bressan2012}. 
The models cover an age range from 1 Myr to several gigayears, and span a wide
range of metallicities.
The models use a Salpeter's initial mass function between 0.15 and 120~M$\odot$.
In order to compare the galactic extinction corrected data with the models,
we construct a grid of synthetic fluxes in all the bands listed in Table~2
by integrating each model SSP spectrum over the corresponding filter responses,
and then dividing by the area of the response curves.
In order to obtain the internal extinction from the SED analysis,
the synthetic broad band fluxes were then reddened using the Cardelli law
\citep{1989Cardelli} for a range of $A_V$ values.

The best fit is obtained by minimizing the merit function $\chi$, calculated as
\begin{equation}\label{pru}
\chi=\frac{1}{N}\sum_{i=1}^{N} \left(\frac{F_{mod}(i)-F_{obs}(i)}{Err(i)}\right)^2
\end{equation}
where $F_{mod}(i)$, $F_{obs}(i)$, and $Err(i)$ are the reddened SSP flux values, the
observed galactic extinction corrected fluxes and observational errors, respectively.
N is the number of bands used in the calculation of $\chi$. The upper-limit 
at the 24~$\mu$m band is not used in our fits.

\begin{figure}
\scalebox{.5}{\includegraphics{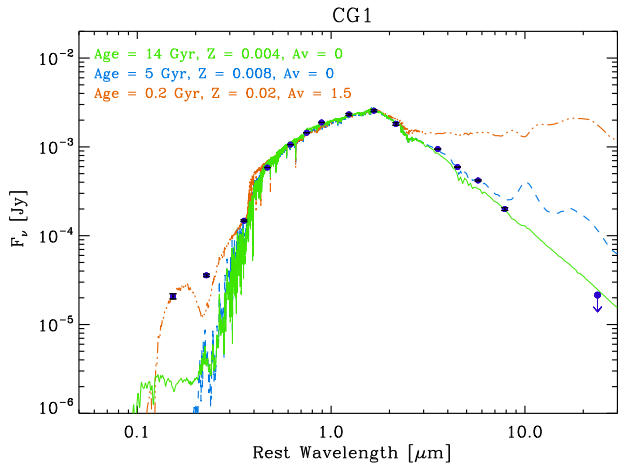}}
\caption{\label{fig:seddeg}
The observed SED of GC1 (from Galex FUV to Spitzer/MIPS 24~$\mu$m)
(blue points) along with SSP models that best fit the SED
(solid lines). None of the SSPs can fit the entire range of SEDs ---
the SSP that fits the UV data (red line) over-produces MIR flux,
the SSP that fits the MIR data (green line) under-produces UV flux.
Use of only the optical and NIR parts of the SED results in age-metallicity
degeneracy (compare the values for the green and blue lines).
The parameters of the models used in
these fits do not produce the blue horizontal branch stars, which
are known to be responsible for the UV emission in Galactic GCs.
}
\end{figure}

\begin{figure}
\scalebox{.5}{\includegraphics{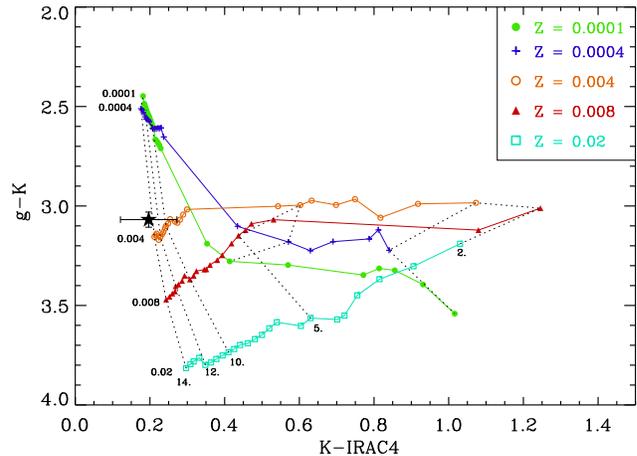}}
\caption{\label{fig:optnircol}
Colour-colour diagram formed using SDSS $g$, 2MASS $K$ and Spitzer 8.0$\mu$m
bands. SSPs are shown for five values of metallicities ranging from Z=0.0001 to
Z=0.02, and various ages between 2--14~Gyr. Dotted (almost vertical) lines join
different metallicity points at fixed ages. The observed colours
(shown by the asterisk) clearly indicate ages $>12$~Gyr, and $Z\la0.004$.
}
\end{figure}

In Figure \ref{fig:seddeg} we show the results of this analysis. The 
points (blue) correspond to the galactic extinction-corrected data, while 
different lines represent different possible solutions that we have in 
our library. The first thing to note is that none of our SSP models can 
simultaneously fit the UV and the MIR part of the spectrum.
Only very young (age$\la 200$~Myr), high metallicity ($Z\ga 0.2$), and
highly reddened ($A_V\ga1.5$~mag) models can reproduce the UV data (dash-dotted 
line). However, the reddening corresponding to $A_V=1.5$ mag in these 
models produces an emission at IR wavelengths very much in excess of the 
MIR data. Notice that the SSP shown in Figure \ref{fig:seddeg} doesn't include 
the reprocessed light corresponding to the absorption of $A_V\ga1.5$~mag, 
inclusion of which will further widen the gap between this model and the 
observed SED at MIR wavelengths. If UV data are not 
considered in the analysis, the best fit corresponds to  an old (formal age$=14$~Gyr), 
relatively metal-poor ($Z=0.004$) and dust-free ($A_V=0.0$~mag) SSP model (green 
solid line). For the sake of completeness, if we restrict the SED to fit only 
the optical to NIR bands, apart from the above solution, equally good fits are 
obtained for a younger (age$=5$~Gyr), but slightly metal-rich ($Z=0.008$) SSP 
with no reddening required, clearly illustrating the effect of the degeneracy 
between age and metallicity (blue dashed line). 
However, even in this case, the model produces an excess emission in the MIR 
part of the spectrum due to the dusty circumstellar envelopes around AGB stars, 
where the typical silicates' emission feature clearly appears at $\sim10~\mu$m. 
The analysis of SSP models that account for dusty circumstellar envelopes show 
that this feature gets stronger at increasing metallicity (and/or at 
intermediate ages), due to the correspondingly higher dust-mass-loss rate of 
the SSPs. On the other hand, the feature vanishes at very low metallicity and/or 
at very old ages. Therefore, the use of the MIR data and, more 
importantly, the upper limit at 24$\mu$m, rules out an age as young as 5~Gyr, and 
favours old ages and close to zero internal extinction. It is the combined 
optical to MIR analysis what ultimately breaks the age-metallicity degeneration 
and favours very old ($>$ 13 Gyr) and moderately metal-poor (Z$\lesssim$ 0.004) 
SSP models. This result is further illustrated in the $g-K$ vs $K-[8.0]$ diagram 
(Figure~\ref{fig:optnircol}), where the observed colours of GC1 (shown by the 
asterisk) indicate an age $>12$~Gyr and metallicity $Z\lesssim0.004$.

However, none of the above models fit the GALEX (and the SDSS-$u$) fluxes. The SSP 
flux in the UV is more than an order of magnitude lower than the observed GALEX 
fluxes, establishing clearly the presence of UV excess. This UV excess is an 
already known issue in massive GCs\citep[e.g.][]{Vink1999, 
DCruz2000, Brown2001, Busso2007}. Blue horizontal branch stars (BHBs) are 
established to be the main reason for the UV excess in GCs.
In canonical models of population synthesis for GCs \citep[e.g.][]{Lee1994},
BHBs naturally appear in old, very metal-poor systems,
whereas metal-rich systems have only red clump stars.
GC1 is not metal-poor, and hence we expect only the red clump, and no UV excess,
especially using the recently downward revised calibration of the
mass-loss rates during the Red Giant Branch \citep{2012MNRAS.419.2077M}.
 However, this problem is not unique to GC1 --- many massive, relatively 
metal-rich galactic GCs are found to have hot BHBs \citep{Rich1997}. 
\cite{Lee1994}
 found these hot BHBs to contain an enhanced amount of helium. 
A second stellar generation with almost equal metallicity, but
He-enriched,  is nowadays the most likely explanation
for the presence of the hot horizontal branch in  galactic GCs 
\citep[e.g.][]{Caloi2007}.

In order to check the possibility that the UV excess could be explained by 
accounting for the presence of a He-rich stellar population, we calculated 
a new set of SSP models for an enhanced value of initial He content. 
In more detail, we use PARSEC code to compute models for the same metallicity of the 
fit ($Z= 0.004$), ages between 9 to 14 Gyr, and an initial He content equal 
to $Y=0.4$, a value that produces entire range of observed $T_{\rm eff}<30000$~K 
for the HB stars \citep{Busso2007}. It is worth noting that these models are 
fully consistent with the ones computed previously by adopting the canonical 
value of $Y = 0.25$, as far as both the physical inputs and numerical 
assumptions are concerned. We re-did the fit, now accounting for data from 
FUV to IRAC4 bands, and where $F_{\rm mod}$ is a combination of the two sets of 
SSPs (one with the canonical Y value, and the other with $Y=0.4$), using
$F_{\rm mod}=(1-f)\times F_{\rm mod,Y=0.25}+f\times F_{\rm mod,Y = 0.4}$, 
where $f$ is the fraction of the He-rich population needed to fit the total SED. 

\begin{figure}
\scalebox{.5}{\includegraphics{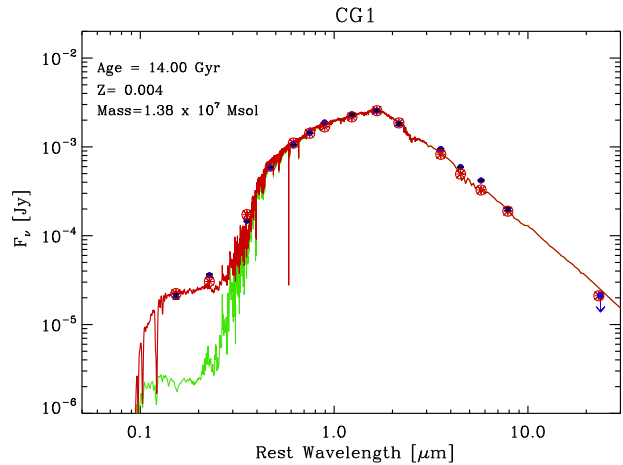}}
\caption{\label{fig:sedhestar}
The observed SED of GC1 (from Galex FUV to Spitzer/MIPS 24~$\mu$m)
(blue points) along with the multiple stellar generation model including
the He-rich population with BHB stars, that best fits the
entire SED (red solid line; 14~Gyr, $Z=0.004$, Y=0.40, $A_V=0.0$~mag).
Green line is the standard model with a single stellar generation
without the BHB stars (see text for more details).
}
\end{figure}

In Figure \ref{fig:sedhestar}, we show the comparison between the observational 
data (blue points),  the best fit (red solid line, and red squares), 
and the previous best fit corresponding to the canonical He abundance (green 
thin solid line). As already discussed, the models computed by assuming a 
canonical He content ($Y=0.25$) are not able to reproduce the UV excess. 
By contrast, the presence of two stellar populations with similar ages 
and metallicities, but markedly distinct initial He content, reproduce the 
entire SED of GC1. The main population corresponding again to an age of 
$\sim$14.0 Gyr, a metallicity of
$Z=0.004$ and the population with canonical He content contributes around 60\% 
to the bolometric luminosity, while the He rich population corresponding to an 
SSP of 13.2 Gyr and a metallicity of $Z=0.004$, contributes 40\% ($f=0.40$) 
to the total luminosity. This result is in agreement with the 
results by \cite{Caloi2007} and  \cite{Busso2007}. They found that significant
fractions, ranging between 35--60\% of a He-rich stellar population, are needed 
in order to explain the morphology and the observational features of the HBs 
of NGC~6441 and NGC~6388. Moreover, by comparing the best models with and 
without a He-rich population (red vs. green lines in 
Fig. \ref{fig:sedhestar}), one can see that the impact of the inclusion of 
the He-rich population is almost negligible at wavelengths longer than the 
$U$-band \citep[see also][]{Girardi2007}.

The SED-inferred metallicity of $Z=0.004$ is in good agreement with that
inferred from our analysis of Lick indices ($Z=0.0043$; see \S4.1),
putting GC1 clearly among the high metallicity GCs. The-SED inferred 
age ($\gtrsim13$~Gyr) and the presence
of second generation of bluer stars are also in good agreement with the 
star formation history inferred using {\sc STARLIGHT}.

Our analysis indicates that the inferred initial mass of GC1 is 
about $1.5\times10^7$ M$_\odot$, 40\% of which are enriched in He content.
The inferred number of He-rich BHB stars (T$_{\rm eff}>7000$~K) is $\sim2523$ 
which produce a bolometric luminosity of about $2.524\times 10^5~\rm L_\odot$. 
\cite{Whitney1994} detected 1957 FUV bright sources in the massive 
galactic cluster $\omega$ Centauri, of which over 30\% are extreme HB stars or 
hot post-AGB stars \citep{DCruz2000}. \cite{Busso2007}, using the star 
counting technique, found 146 and 218 BHB stars in NGC 6441 and NGC 6388, 
respectively, two massive ($\sim 10^6$ M$_\odot$), old ($\sim11$--13~Gyr) 
and metal-rich ($[Fe/H]\thickapprox -0.5$) bulge globular clusters. 
\cite{2009Ma} used  the broad-band (FUV to NIR) SED fitting technique in 
G1 in M31, and found also a UV excess which would correspond to 
$\sim165~\rm L_\odot$ FUV-bright, hot, extreme HB stars, which is more 
than 3 order of magnitude lower than the results we got for GC1, even though 
M31-G1 has a similar mass as GC1. However, we note that M31-G1 is metal-poor 
by more than a factor of 2 as compared to GC1. On the other hand,
note that GC1, NGC~6441, NGC~6388 and even $\omega$ Centauri are massive and 
metal-rich globular clusters ($Z\gtrsim 0.004$), and only a He-rich old 
stellar population could give a significantly higher number of hot, evolved HB 
stars, as compared to the He-poor counterparts of the same age and lower metallicity
\citep[e.g.][]{Lee1990,Lee2001,Yi1999,Caloi2007,Busso2007}. The presence of 
He-rich stellar populations have also been proposed by \cite{Kaviraj2007b} to 
explain the UV emission of UV-bright clusters in M87. As far as the authors 
know, GC1 could be the cluster with the largest number of He-rich BHB stars. 
 
We would like to point out, and it is easy to see in Fig. \ref{fig:sedhestar}, 
that the best fit (red solid line) model under-produces emission at the NUV 
with respect to the observed data, most likely indicating that the BHB stars
in our models are too hot. It is well established that the bluest HB stars 
are cooler at lower He enhancement values 
\citep[e.g.][]{Raimondo2002,Dantona2005,Moehler2006,Caloi2007,Busso2007}.
Thus, it is likely that the He enhancement in GC1 is not as high as $Y=0.4$,
but $\sim0.35$. Determining the exact value of $Y$
is beyond the scope of this work, as it also depends on the 
mass-loss efficiency during the red giant evolution.
 
The presence of He-enriched stellar population would imply that GC1 must 
have had at least two episodes of star formation with the second generation 
of stars polluted by material ejected from the first generation of stars. 
The stars responsible for the pollution could be type II supernovae, 
rotating massive stars or massive-AGB stars \citep[e.g.][]{Dantona2002, Renzini2008}. 
The nature of the progenitor and how the ejected material can remain 
inside the potential well will be discussed in \S5.
 
Given that the presence of a small fraction of He-enriched stars can give rise 
to an UV excess, it is not advisable to use UV fluxes while fitting SEDs to 
obtain age and metallicity of old simple populations. Ignoring the He-enriched 
stars in clusters with intrinsic UV excess would lead to overestimates of 
ages such as found in \cite{Kaviraj2007a} and \cite{2009Ma}. 

\section{DISCUSSION AND CONCLUSIONS}

\subsection{Photometric mass}

The derived photometric mass of $1.5\times10^7$~\msol\ corresponds to the mass 
at birth of the cluster using the Salpeter IMF (see \S4.3).
At the present age of $\sim13.5$~Gyr, the cluster still contains 65\% of the 
initial mass (51\% in living stars, and 14\% in stellar remnants). Thus, the present 
mass of the cluster with the Salpeter IMF is $1.0\times10^7$~\msol. 
With the \cite{Kroupa2001} IMF, the present mass would be $6.3\times10^6$~\msol.
The derived mass is comparable to the photometric mass of M31-G1 (see Table~1), 
whereas it is $\sim4$ times more than that of $\omega$~Cen, the brightest GC in the 
Milky Way. 
In this section, we analyse whether this cluster shares properties that are
established to be characteristics of massive GCs, and address
the issue of its origin.

\subsection{Dynamical state}

The relatively high concentration index of $c=\log(r_t/r_c)\sim1.88$ suggests 
that the cluster is in a post-core collapse stage, where binaries at the 
centre of the cluster provide a source of energy to halt the collapse. 
The GC1 is expected to harbour a number of close low-mass binaries,
which are expected to emit X-rays.
The brightest GCs in galaxies are known to be X-ray emitters
(e.g. \cite{Kundu2002} in NGC 4472; \cite{Fabbiano2010} in NGC 4278),
where the low-mass X-ray binaries are responsible for the X-ray emission.
The X-ray missions ROSAT and Chandra, both have 
detected X-ray emission from GC1 \citep{Immler2001, Swartz2003},
with the latter
reporting an X-ray luminosity of $5.1\pm1.1\times 10^{37}$erg~s$^{-1}$.
The observed luminosity of GC1 corresponds to a system of a binary where
the donor is an evolved giant star \citep{Revnivtsev2011}.

The cluster is at a projected distance of only 3.0~kpc from the nucleus of the
galaxy. We now calculate the expected tidal radius of GC1, using the
relation given by \cite{Spi87}: \\
$R_{\rm t} = \left({M_{\rm C}\over{2M_{\rm G}}}\right)^{1/3} R_{\rm G}$,\\
where $R_{\rm G}=3.0$~kpc is the galactocentric radius of GC1, 
$M_{\rm C}=1.5\times10^{7}$~\msol\ is the mass of GC1, and
$M_{\rm G}$ is the mass of the parent galaxy within the radius $R_{\rm G}$.
\cite{2010Nantais} estimate
a mass of $0.88\times10^{11}$~\msol\ within a galactocentric radius of 3.82~kpc.
Substituting these values, we get a tidal radius of 116~pc. 
The observed $r_t=93$~pc in the $F814$-band is 80\% of this value.  
Thus, the dynamical evolution of GC1 is not being affected by the tidal forces of the
parent galaxy, unless the cluster is in a highly eccentric orbit and that
it had a peri-centre radius as small as 2.0~kpc. The fact that the observed
radial velocity of GC1 is consistent with that expected at the present radius,
possibly rules out the object being in a highly eccentric orbit.

In a recent work, \cite{Gieles2011} divided the Galactic GCs into
two categories --- the expansion-dominated and evaporation-dominated.
GCs in the first category are massive ones that are evolving without the tidal
effects of the parent galaxy, resulting in the increase of their half-mass radius 
with time. The GCs in the second category are tidally limited, resulting in 
evaporation, and subsequent contraction. The observed mass of GC1 clearly puts
it in the first category. If we extrapolate the $r_{\rm h}$-Mass relation that
\cite{Gieles2011} obtained for the Galactic GCs to the mass of GC1, we obtain an 
$r_{\rm h}=1.26$~pc, which is around 4 times smaller as compared
to the observed value. Thus, GC1 is in the expansion-dominated phase.

\subsection{Metallicity and $\alpha$-enrichment} 

Colour distribution of GCs shows bimodality, which is principally due to a 
metallicity difference between two old populations \citep{2006Brodie}, with the
metal-poor ($[Fe/H] < -1$) GCs being relatively bluer than their metal-rich 
counterparts. GC1 is metal-rich $[Fe/H] = -0.60\pm0.10$, and 
moderately $\alpha$-enriched ([$\alpha/{\rm Fe}]\sim0.2\pm0.05$). 
Star formation episodes extending for more than $\gtrsim1$~Gyr are not 
expected to show $\alpha$ enrichment, and hence the observed value of 
[$\alpha/{\rm Fe}$] rules out extended period of star formation. Thus, star 
formation and metal enrichment in this cluster should have happened over this 
short time-scale. This implies that the cluster was efficient in retaining 
all or most of metals ejected in the initial burst.
Metals are expelled from stars in the form of high velocity winds
of high mass stars, through explosion of SNII, and through the winds of
AGB stars. With the inferred photometric mass of $1.0\times10^7$~\msol, and
a concentration parameter of $\sim1.88$, we estimate an escape velocity at the tidal 
radius of 146~km\,s$^{-1}$ at present, using the expression given by 
\cite{Georgiev2009}. Given that the initial cluster mass 
is expected to be higher, and the $r_h$ lower than the presently observed
values, the escape velocities during the first gigayear of star formation
would have been higher than this value. Terminal velocity of winds in AGBs 
of stars of metallicities $[Fe/H] = -0.60$ is expected to be around 
50~km\,s$^{-1}$, whereas the velocity of winds from high-mass stars, and supernova 
ejecta would be much higher than the escape velocities. 
However, the cluster potential was deep enough to trap metals from AGB winds, 
resulting in enrichment of the interstellar medium before the star formation
ceased. The potential well, however, was not deep enough to trap all the 
metals generated in the cluster. 

Trapping of metal-enriched gas also leads to the formation of helium enriched
second generation of stars, that at present show up as blue HB stars.
The SED of the cluster, especially the flux in the Galex bands, clearly
suggests an extended blue HB, consisting of $\sim2500$ stars.

\subsection{Is GC1 the nucleus of a dissolved dwarf galaxy?}

\begin{figure}
\scalebox{1.0}{\includegraphics{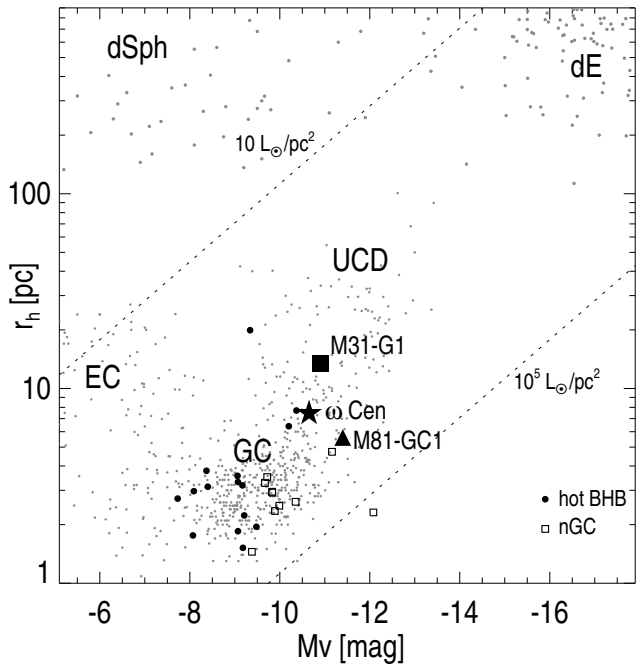}}
\caption{\label{fig:sizelum}
The location of GC1 in the size-luminosity plane compared to for various 
kinds of spherodal systems (dSph: dwarf spherodals, dE: dwarf ellipticals,
UCD: ultra compact dwarfs, EC: extended or fuzzy clusters, nGC: nuclei of
dwarf galaxies, E-BHB: Galactic GCs with extended BHB).
Data for galactic and extragalactic GCs are denoted by GC. The two well-known
massive clusters, $\omega$~Cen and M31-G1 are identified. The lines of constant
mean surface brightness of $10^5$ and 10~L\sun/pc$^2$ are shown by the two 
diagonal lines. During stripping of dwarf galaxies, their nuclei would fade
and expand, resulting in decrease of their surface brightness. 
Like $\omega$~Cen and M31-G1, a high surface brightness nGC could be the 
progenitor of GC1.
}
\end{figure}

GC1 is clearly one of the most massive clusters in the local Universe. 
Ever since the success of the numerical simulations
of \cite{Bekki2003} in explaining the observed properties of $\omega$~Cen, 
massive GCs are often considered as nuclei of stripped dwarf galaxies.
We here discuss whether GC1 also fits into this picture.
Kormendy's classical work \citep{1985Kormendy} led to the use of
observational planes formed from two or more of the following four 
quantities --- 
central or mean surface brightness, core or half-light radius, 
total absolute magnitude and central 
velocity dispersions --- to address the 
inter-relation between different spheroidal systems %issue of origin of GCs
\citep[e.g.][]{Boselli2008, 2010Vanden, Taylor2010}.

\cite{Georgiev2009} used the $r_{\rm h}$ vs $M_v$ diagram to address
the origin of GCs that have hot BHBs. More recently, 
\cite{2011Brodie} and \cite{2013Forbes} have used this diagram to illustrate 
the continuity of properties of different spheroidal systems. 
In Figure~\ref{fig:sizelum}, we show the $r_{\rm h}$ vs 
$M_v$ diagram, where data for GC1 are plotted along with those for other
spheroidal systems. Data for galactic and extragalactic GCs, extended 
clusters (EC; also known as Fuzzy Clusters), UCDs and cores of dwarf 
spheroidals (dSph) and dwarf ellipticals (dE) are taken from \cite{2011Brodie}. 
Data for nucleated dwarf galaxies (nGC; also known as nuclear GCs), 
and galactic GCs that have hot BHBs from \citet{Georgiev2009}.
Data for M31-G1 and $\omega$~Cen were taken from the sources listed in 
Table~\ref{Tab:SuperClusters}.
The diagonal lines show the locus of constant mean surface density of $10^5$
and 10~L\sun/pc$^2$, within the half-light radius.
During stripping of a dwarf galaxy, its nucleus is expected to experience 
expansion and also fade in intensity \citep{Bekki2003}. This would 
reduce the surface brightness, moving the points roughly
along a direction perpendicular to the constant surface brightness lines.

With a mean surface density of $1.5\times10^4$~L\sun/pc$^2$,
GC1 is among the highest surface density objects, especially among the
luminous objects. The classes of objects that are more massive than GC1 
are cores of dEs, UCDs and nGCs.
The progenitor candidate should have higher surface density than the
presently observed value for GC1 to account for the expansion and fading
associated with stripping. The high surface brightness nucleated dwarf 
galaxies are the only objects satisfying this criterion, and 
hence are the most likely progenitors of GC1.
It is interesting to note that
the observed ellipticity of GC1 ($\epsilon=0.12$), is 
almost identical to the mean ellipticity of nuclei of dwarf galaxies 
($<\epsilon>=0.11$; \cite{Georgiev2009}). \cite{Georgiev2009} propose 
nuclei of dwarf galaxies as progenitors of GCs with hot BHBs, 
a property shared by GC1. Thus, all the observed evidence points 
towards a dwarf galaxy nuclear origin for GC1. 

It is most likely that the nucleated dwarf galaxy that was once upon a time
the progenitor of GC1 was intact for at least the first 1 Gyr, helping in its metal-enrichment.
Subsequently, as the dwarf galaxy started accreting onto M81, the tidal forces
dissolved the galaxy, leaving behind the compact nucleus indistinguishable in
appearance from a classical GC. 

\section{Summary}

We investigate the nature of the brightest GC in M81 by carrying out a detailed
analysis of multi-band photometric and optical spectroscopic data. 
We establish that the cluster is old (age$\ga13$~Gyr) and metal-rich ($[Fe/H]=-0.60\pm0.10$).
The UV excess suggests the presence of $\sim2500$ hot blue horizontal branch stars, a
characteristic common in many metal-rich GCs. The cluster is bluer in its core,
suggesting that the hot BHB stars are concentrated at the centre of the
cluster. The radial profile of the cluster can be fitted very well with a 
King profile of a core radius $r_c=1.2$~pc, and a logarithmic concentration index of 1.88. 
The low $r_c$ and high concentration index suggest that the cluster is in a 
post core-collapse phase. All the observed properties of GC1 support the idea
that it could be the left-over nucleus of a dwarf galaxy that has been
dissolved during its accretion onto M81 in the early epoch of its formation.

\section{ACKNOWLEDGMENTS}
We would like to thank the Hubble Heritage Team at the Space Telescope Science
Institute for making the M81 images publicly available. 
We also thank the anonymous referee for many useful comments that have led 
to an improvement of the original manuscript.
This work is partly supported by CONACyT (Mexico) research grants CB-2010-
01-155142-G3 (PI:YDM), CB-2011-01-167281-F3 (PI:DRG) and CB-2012-183013 (PI:OV).

\label{lastpage}
\end{document}